\documentclass[prl,aps,twocolumn,floatfix,superscriptaddress,a4paper]{revtex4}

\usepackage{amsmath,amssymb,amsfonts}
\usepackage{bm}\let\vec\bm
\usepackage{graphicx}

\usepackage{color}

\begin{document}

\title{Dynamical properties of the one-dimensional spin-$1/2$ Bose-Hubbard model near Mott-insulator to ferromagnetic liquid transition }

\author{M. B. Zvonarev}
\affiliation{DPMC-MaNEP, University of Geneva, 24 quai Ernest-Ansermet, 1211 Geneva 4, Switzerland}

\author{V. V. Cheianov}
\affiliation{Physics Department, Lancaster University, Lancaster, LA1 4YB, UK}

\author{T. Giamarchi}
\affiliation{DPMC-MaNEP, University of Geneva, 24 quai Ernest-Ansermet, 1211 Geneva 4, Switzerland}

\begin{abstract}

We investigate the dynamics of the one-dimensional strongly repulsive spin-$1/2$ Bose-Hubbard model for filling $\nu\le1.$ While at $\nu=1$ the system is a Hubbard-Mott insulator exhibiting dynamical properties of the Heisenberg ferromagnet, at $\nu<1$ it is a ferromagnetic liquid with complex spin dynamics. We find that close to the insulator-liquid transition the system admits for a complete separation of spin and density degrees of freedom valid at {\it all} energy and momentum scales within the $t-J$ approximation. This allows us to derive the propagator of transverse spin waves and the shape of the magnon peak in the dynamic spin structure factor.

\end{abstract}

\date{\today}
\maketitle

Collective dynamics of interacting quantum particles in a periodic potential is a paradigmatic problem of many-body physics. The simplest model incorporating the competition between the interaction and kinetic energy on a lattice is the Hubbard model~\cite{hubbard_insulator_transition}. For fermions, this model has been the cornerstone of the theory of strongly correlated systems. In particular, it describes the transition from a Mott insulator at a commensurate filling to a metallic state when the system is doped. The nature and properties of the weakly doped state remain the subject of fierce debate, in which an interplay between the spin and charge dynamics occupies a central place. It is this interplay that is believed to lead to a rich phase diagram with putative phases ranging from a ``normal'' Fermi liquid state with heavy holes propagating in an antiferromagnetic background to a d-wave superconducting state arising from a spin liquid. The bosonic version of the Hubbard model, the Bose-Hubbard (BH) model~\cite{fisher_boson_loc}, also has a wide range of applications, from granular superconductors to Josephson junction arrays. Recent experimental progress in ultracold atomic gases further increased the interest to this model~\cite{bloch_QG_review}.

Like its fermionic sibling, the BH model exhibits a Mott-insulator to liquid transition. This transition has already been observed experimentally in the ultracold one-component (``spinless'') boson gases~\cite{greiner_mott_bec, spielman_2D_MI07, stoferle_tonks_optical} and is likely to be seen in multicomponent (``spinful'') gases soon. A principal difference between the fermionic and bosonic case comes from their different magnetic orders: while the fermionic ground state is antiferromagnetic, the bosons have a tendency to ferromagnetism. This difference should have important consequences for the Mott-insulator to liquid transition. The essentially non-perturbative nature of this transition makes its theoretical analysis a non-trivial task. Therefore, the particular case of one special dimension (1D), where non-perturbative methods are well developed is of special importance. For both fermionic Hubbard and spinless BH models in 1D the phase obtained by doping the Mott insulator is the Luttinger Luquid (LL), whose properties are well established~\cite{giamarchi_book_1d}. However, due to ferromagnetism, the doped state of the BH model with spin cannot be a LL. For a continuous system it was demonstrated very recently~\cite{zvonarev_ferrobosons07} that ferromagnetism leads to a new phase, named ferromagnetic liquid, with very unusual properties~\cite{zvonarev_ferrobosons07, akhanjee_ferrobosons07, matveev_isospin_bosons08, kamenev_spinor_bosons08}. What is the analogue of that phase for a doped BH insulator, particularly in the vicinity of the insulator to liquid transition, is an open question. The analysis of this problem is especially deserving since spin-resolved measurements in a 1D boson quantum gas were already reported~\cite{widera_spinLL_2comp08}.


In this Letter we consider the two-component (spin-$1/2$) 1D Bose-Hubbard model with strong on-site repulsion. We find that in the vicinity of the liquid to Mott-insulator transition the system admits for a complete spin-charge separation at {\it all} energy and momentum scales within the $t-J$ approximation. This result makes possible explicit calculation of dynamic correlation functions of the system. In particular, we calculate the Green's function of transverse spin excitation (magnon) and derive the shape of the magnon peak in the dynamic spin structure factor.

The Hamiltonian of the repulsive ($U>0$) spin $1/2$ BH model on
a 1D lattice with $M$ sites is
\begin{equation}
H= -t_h\sum_{\substack{j=1\\\alpha=\uparrow,\downarrow}}^M (b^\dagger_{\alpha j}b_{\alpha j+1}+ \mathrm{h.c.})\\
+U \sum_{j=1}^M \varrho_j(\varrho_j-1),  \label{Hbose1}
\end{equation}
In this expressions $b^\dagger_{\alpha j}, b_{\alpha j}$ and
$\varrho_{j}=b^\dagger_{\downarrow j} b_{\downarrow
j}+b^\dagger_{\uparrow j} b_{\uparrow j}$ are the creation,
annihilation and the particle number operators on the $j$-th
lattice site, respectively. For a given particle number $N$ one
defines the filling factor $\nu=N/M.$

We derive an effective Hamiltonian valid for large on-site repulsion, $U\gg t_h,$ from Eq.~\eqref{Hbose1} in analogy with the $t$-$J$ approximation for the Hubbard model~\cite{essler_book_1DHubbard}. This is done by treating the potential term in Eq.~\eqref{Hbose1} as the leading order Hamiltonian, the kinetic term as a perturbation, and expanding over this perturbation to the second order. It is convenient to write the resulting $t$-$J$ Hamiltonian using the operators  $c^\dagger,$ $c$ and $\ell,$ which have been  previously used for the analysis of the collective dynamics of spin and charge excitations in fermion~\cite{ogata_inf} and boson~\cite{akhanjee_ferrobosons07, matveev_isospin_bosons08} systems. We have $H_{\mathrm{eff}}= H_c+ H_s, $ where
\begin{equation}
H_c= -t_h\sum_{j=1}^M (c^\dagger_{j}c_{j+1}+ \mathrm{h.c.}), \label{Kbose2}
\end{equation}
and
\begin{multline}
H_s= \frac{t_h^2}{2U}\sum_{j=1}^M \left(c_{j}^\dagger c_{j-1}^\dagger
c_{j+1} c_{j} + c_{j+2}^\dagger c_{j+1}^\dagger c_{j+1} c_{j} \right. \\ \left. +2 c_{j+1}^\dagger c_{j}^\dagger c_{j+1} c_{j} \right)
\times [\vec \ell(\mathcal{N}_j)+\vec \ell(\mathcal{N}_j+1)]^2 .
\label{ham1}
\end{multline}
In these expressions $c^\dagger_j$ ($c_j$) are spinless fermion creation (annihilation) operators, $\varrho_j=c^\dagger_j c_j,$ and $\mathcal{N}_j= \sum_{n=1}^j \varrho_n$ counts the number of particles between the first and $j$-th lattice site. The operator $\ell_s(\mathcal{N}_j),$ $s=x,y,z$ represents the spin state of the nearest particle to the left of the site $j$ and is defined by
\begin{equation}
\ell_s(\mathcal{N}_j)= \sum_{m=1}^N \frac{\ell_s(m)}{2\pi} \int_0^{2\pi} d\lambda\, e^{i\lambda\left(m-\mathcal{N}_j\right)}, \label{l_-N}
\end{equation}
where $\ell_s(m)$ are local spin-$1/2$ operators on a lattice, $[\ell_x(m),\ell_y(n)]= i\delta_{mn}\ell_z(n).$ Note that in the sector with maximal value of the total spin Eq.~\eqref{ham1} gives the $t-J$ approximation of the spinless BH model, cf.\ Eq.~(4) of Ref.~\cite{cazalilla_hcbose_gases}.

While $H_c$ generates the time evolution of free spinless fermions, spin and charge (density) degrees of freedom are coupled in $H_s$ non-trivially. An advantage of the representation~\eqref{ham1} is that in the vicinity of the Mott phase, $1-\nu\ll1,$ which we focus on in our Letter, one can neglect quantum fluctuations of $c$-dependent operators in $H_s.$ As a result one gets the Hamiltonian of the Heisenberg ferromagnet
\begin{equation}
H_{s}= -2J \sum_{j=1}^N \left[\vec\ell(j) \vec\ell(j+1)- \frac14\right], \label{Hheis}
\end{equation}
with the exchange constant $J=t_h^2/U.$ We shall see that already within the approximation~\eqref{Hheis} the dynamics of the system is non-trivial. The analysis of perturbative corrections in $1-\nu$ to Eq.~\eqref{Hheis} remains an interesting open question.

We now proceed to the investigation of spin dynamics above the fully polarized ground state $|\Uparrow\rangle$ of the Hamiltonian~\eqref{Hbose1}. Like the ferromagnetic Bose gas~\cite{zvonarev_ferrobosons07}, a non-trivial part of this dynamics is encoded in the correlation function of transverse spin excitations
\begin{equation}
G_\perp(j,t)= \langle\Uparrow|s_+(j,t)s_-(0,0)|\Uparrow\rangle, \label{Gperpdef}
\end{equation}
where $s_+(j)= b^\dagger_{\uparrow j} b_{\downarrow j}$ and $s_-(j)=[s_+(j)]^\dagger.$ Since $G_\perp(j,t)=G_\perp(-j,t)$ and $G^*_\perp(j,t)=G_\perp(-j,-t)$ we consider $j,t\ge 0.$ We focus on the regime of weak hole doping, $\mu,$ and strong on-site repulsion:
\begin{equation}
\mu=1-\nu\ll 1 \quad \text{and} \quad Jt_h^{-1}=t_h U^{-1}\ll 1. \label{conditions}
\end{equation}
Length and time scales in our problem in the regime~\eqref{conditions} are set up by the parameters $q_F$ and $t_F,$ respectively:
\begin{equation}
q_F=\pi \mu, \qquad t_F= E_F^{-1}, \qquad E_F= q_F^2t_h. \label{cond2}
\end{equation}
Combining Eqs.~\eqref{conditions} and~\eqref{cond2} we get three dimensionless small parameters: $q_F,$ $J/t_h,$ and $E_F/t_h.$

To calculate $G_\perp$ using Eqs.~\eqref{Kbose2} and \eqref{Hheis}, we express the operators  $s_\pm$  through $c^\dagger,$ $c,$ and $\ell_\pm=\ell_x\pm i\ell_y$:
\begin{equation}
s_-(j)= \varrho_j \ell_-(\mathcal{N}_j) \label{sdec}
\end{equation}
and factorize the ground state
$|\Uparrow\rangle= |\Uparrow\rangle_H \otimes |\mathrm{FS}\rangle, $
where $|\Uparrow\rangle_H $ is the fully polarized ground state of $H_s,$ Eq.~\eqref{Hheis}, and $|\mathrm{FS}\rangle$ is the ground state of $H_c,$ Eq.~\eqref{Kbose2}, at a given~$\nu$. For Eq.~\eqref{Gperpdef} this gives
\begin{equation}
G_\perp(j,t)= \int_{-\pi}^\pi d\lambda\, G_H(\lambda,t)
D_\nu(\lambda; j,t ). \label{GperpH1}
\end{equation}
Here $G_H(\lambda,t)=\exp\{-iE_s(\lambda)\}/2\pi$ is the space Fourier transform of the transverse spin-spin Green's function $\phantom|_H\langle \Uparrow \vert \ell_+(n,t)\ell_-(0,0)|\Uparrow\rangle_H$ of the Heisenberg ferromagnet~\eqref{Hheis}, and $E_s(\lambda)=2Jt(1-\cos\lambda)$ is the dispersion of a single magnon in that model. The fluctuations of density enter Eq.~\eqref{GperpH1} through the Green's function~\cite{rhoapp}
\begin{equation}
D_\nu (\lambda; j,t)=\langle\mathrm{FS} \vert
e^{-i\lambda\mathcal{N}_j(t)}
e^{i\lambda\mathcal{N}_0(0)} \vert \mathrm{FS} \rangle. \label{D}
\end{equation}
We stress that though $G_H$ ($D_\nu$) contains collective spin (density) degrees of freedom only, spin and density are coupled to each other in $G_\perp$ through the integration over $\lambda$ in Eq.~\eqref{GperpH1}.

\begin{figure}
\includegraphics[width=8 cm]{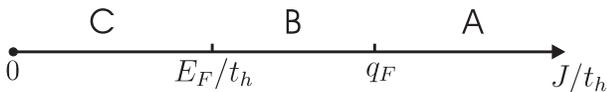}
\caption{Shown are the parametric regimes (A), (B), and (C) corresponding to different relative values of the parameters $q_F=\pi\mu,$ $J/t_h,$ and $E_F/t_h$.
}
\label{fig:regions}
\end{figure}

There are two ways to analyze Eq.~\eqref{D}. Like other free fermion Green's functions~\cite{izergin_impenetrable_fermions}, it can be represented as a Fredholm determinant~\cite{zvonarev_string09}. The large space, $q_F j> 1,$ or time, $t/t_F> 1,$ asymptotics of this determinant can be found using the results of Ref.~\cite{cheianov_spin_decoherent_long}, though the calculations are rather involved. Alternatively, large space-time asymptotics of Eq.~\eqref{D} can be calculated using the LL formalism. There the operator $\mathcal{N}_j(t)$ is given by $\mathcal{N}_j(t) = \nu j -\pi^{-1} \partial_x\phi(x,t),$ where $\phi$ is a free bosonic field and $x=a j,$ $a$ being the lattice spacing~\cite{giamLL}. Thus
\begin{equation}
D_\nu(\lambda;j,t)\simeq \exp\left\{-i\lambda \nu j -\frac{\lambda^2}{4\pi^2} \ln\frac{|j^2-v_F^2 t^2|}{v_F^2 t_c^2}\right\}. \label{av1}
\end{equation}
Here $v_F=2t_h q_F$ is the Fermi velocity of the holes, which coincides with the sound velocity of the system. The cutoff parameter $t_c\sim t_F.$ Its explicit value cannot be determined within the LL formalism, while the analysis of the Fredholm determinant gives $t_c= e^{-1-\gamma-\ln 4} t_F \approx 5.16\times 10^{-2} t_F$, where $\gamma=0.577\ldots$ is the Euler-Mascheroni constant~\cite{zvonarev_bose-hubbard_long08}. Having the result~\eqref{av1}, we can investigate different parametric regimes shown in Fig.~\ref{fig:regions}.

{\it(A):} $J/t_h> q_F.$ The doping is so small that during a large time $t_*$ the spin excitation evolves like in the undoped system:
\begin{equation}
G_\perp(j,t)= e^{-i\frac\pi2 j} e^{2iJt} \mathcal{J}_j(2Jt), \qquad t<t_*,
\label{Heis}
\end{equation}
where $\mathcal{J}_j$ is the Bessel function of the first kind. To estimate $t_*$ note that Eq.~\eqref{Heis} can be obtained by letting $\nu=1$ in Eq.~\eqref{D} which implies $D_{\nu=1}(\lambda;j, t)=e^{i\lambda j}$.
Such an approximation is valid as long as $q_F j < 1$ and $t < t_F.$ As $j\to\infty$ one has $\mathcal{J}_j\sim e^{-j\ln j}$ starting from $j \sim 2Jt$ and due to the exponential decay possible deviations of $G_\perp$ from what is predicted by Eq.~\eqref{Heis} become immaterial for larger $j.$ One thus arrives at the condition $2Jq_Ft < 1,$ which implies $t_*=1/2Jq_F<t_F.$

For $t>t_F$ one can use Eq.~\eqref{av1} and the integral in Eq.~\eqref{GperpH1} is dominated by the saddle point
$ \lambda_s=\arcsin\left(- j/2Jt -i0 \right)$.
Therefore
\begin{equation}
G_\perp(j,t) \simeq \sqrt{\frac{\pi}{iJt\cos\lambda_s} } G_H(\lambda_s,t)D_\nu(\lambda_s;j,t)
\label{Gescape}
\end{equation}
with $D_\nu$ given by Eq.~\eqref{av1}. One can see that $G_\perp$ oscillates as a function of $j$ for $j<2Jt$ and decays as $e^{-j\ln j}$ for $j>2Jt.$ The divergence of Eq.~\eqref{Gescape} as $j\to v_F t$, customarily called light-cone singularity, indicates what fraction of a collective excitation propagates with the velocity of sound.

{\it (B):} $E_F/t_h<J/t_h<q_F.$ Here the result \eqref{Heis} is valid for  $t<t_F$ and it crosses over to Eq.~\eqref{Gescape} for $t>t_F$. The main difference from the case (A) is the absence of the light-cone singularity. Indeed, the exponential decay of Eq.~\eqref{Gescape} due to the imaginary part of $\lambda_s$ starts before $j$ reaches the light cone, $v_Ft.$ Note that one can use $D_\nu(\lambda_s;j,t) \simeq e^{-i\lambda_s j} D_\nu(\lambda_s;0, t)$ in Eq.~\eqref{Gescape} in this regime.

{\it(C):} $J/t_h<E_F/t_h.$ Here for $t<t_F$ the spin excitation does not propagate at all, $G_\perp(j,t)=\delta_{0j}.$ For $t>t_F$ and  $2\pi^2 Jt < \ln (t/t_c)$ (the latter inequality could be approximately written as $t<J^{-1}$) the integral in~Eq.~\eqref{GperpH1} is dominated by the region $\vert\lambda\vert<2\pi/\sqrt{\ln (t/t_c)}$ and the result of the integration is~\cite{SU2}
\begin{equation}
G_\perp(j,t) \simeq \sqrt{\frac{\pi}{2 \ln (t/t_c)}}
\exp\left\{-\frac{(\pi j)^2}{2\ln(t/t_c)}\right\} . \label{Gregion1}
\end{equation}
For this calculation we replaced $|j-v_Ft|$ by $v_F t$ in Eq.~\eqref{av1}: keeping the light cone singularity in Eq.~\eqref{Gregion1} would be unphysical since $G_\perp$ decays as a gaussian at distances $j \sim \sqrt {\ln t/t_c}\ll v_F t.$ The physical interpretation of Eq.~\eqref{Gregion1} is that the propagation of spin excitation is caused by the fluctuations of the position of the particle carrying this spin rather than by exchange processes. The same propagation mechanism takes place in the continuum limit in the trapped regime, cf.\ Eq.~(6) of Ref.~\cite{zvonarev_ferrobosons07}.
Finally, for $t>t_F$ and $2\pi^2 Jt > \ln (t/t_c)$ the asymptotics of $G_\perp$ is given by Eq.~\eqref{Gescape} and the analysis carried out for the case (B) applies. In particular, $G_\perp\sim e^{-j\ln j}$ as $j\to\infty,$ which is pronouncedly different from the gaussian decay of $G_\perp$ in the continuum limit, cf.\ Eq.~(16) of Ref.~\cite{zvonarev_ferrobosons07}. This difference stems from different propagation mechanisms for the spin excitations in the $\nu\ll1$ and $1-\nu\ll1$ regimes: while in the former both longitudinal and transverse spin excitations are carried by particles, in the latter longitudinal spin (that is, density) excitations are carried by holes and transverse spin excitations by particles.

The rest of this Letter is devoted to the analysis of the spectral function $A(k,\omega)= \sum_f \delta(\omega-E_f(k))|\langle f,k|s_-(k)|\Uparrow\rangle|^2,$ where $H|f,k\rangle= E_f(k)|f,k\rangle$ and $f$ enumerates all the states with a given momentum~$k.$ Using Eq.~\eqref{GperpH1} and the identity $D_\nu(\lambda; j,t)=e^{-i \lambda j} D_{\mu}(-\lambda;j,t)$ following from particle-hole symmetry we write
\begin{equation}
A(k,\omega)= \int_{-\pi}^\pi \frac{d\lambda}{2\pi} \tilde D_{\mu} (\lambda;k-\lambda,\omega-E_s(\lambda)), \label{Afourier}
\end{equation}
where $\tilde D_\mu$ is the Fourier transform of $D_\mu(\lambda;j,t)$ with respect to $j$ and $t$ and $E_s(\lambda)$ is defined below Eq.~\eqref{GperpH1}. For $k$ and $\omega\to0$ Eq.~\eqref{av1} implies
\begin{equation}
\tilde D_\mu(\lambda;k,\omega)\sim e^{-\omega t_c}\theta(\omega-2\pi|p|) [\omega^2-(2\pi p)^2]^{\alpha(\lambda)}, \label{Dfourier}
\end{equation}
where $p=k+\lambda\mu$ and $\alpha(\lambda)=-1+(\lambda/2\pi)^2$. Substituting Eq.~\eqref{Dfourier} into \eqref{Afourier} we arrive at
\begin{equation}
A(k,\omega) \sim \theta(\omega-E_s(k/\nu)) [\omega-E_s(k/\nu)]^{\Delta(k)} \label{Akomegahresh}
\end{equation}
valid for $\omega$ close to $E_s(k/\nu)$ and $-\pi< |k|< \pi.$ The threshold exponent $\Delta(k)$ in Eq.~\eqref{Akomegahresh} is
\begin{equation}
\Delta(k)= -1+ \frac{k^2}{2\pi^2}. \label{Delta}
\end{equation}
Equations~\eqref{Akomegahresh} and \eqref{Delta} constitute one of the main results of this Letter, so let us discuss them in more detail. The threshold exponents of the spectral functions were calculated for several classes of 1D interacting models, mostly using quite involved techniques~\cite{cheianov_structfactor_integrable08}. Probably, the closest to our work are the studies of $A(k,\omega)$ in 1D ferromagnetic Bose gas, the model defined by taking the continuum limit of Eq.~\eqref{Hbose1}. In this model $\Delta(k),$ found in Refs.~\cite{zvonarev_ferrobosons07, matveev_isospin_bosons08, kamenev_spinor_bosons08}, is also given by Eq.~\eqref{Delta}, but with different coefficient in front of $k^2$ term. Moreover, according to Ref.~\cite{kamenev_spinor_bosons08}, any Galilean-invariant system of Bose particles with ferromagnetic ground state should have $\Delta(k)$ given by Eq.~\eqref{Delta}. We see that the BH model, which is not Galilean-invariant, still has the threshold exponent~\eqref{Delta} at $1-\nu\ll1$. Whether this is an accident, or a manifestation of some fundamental principle other than Galilean invariance, is an open question. A deeper understanding of this issue could be obtained by calculating the threshold exponent for the BH model at arbitrary filling, which is not yet done.

\begin{figure}
\includegraphics[width=8 cm]{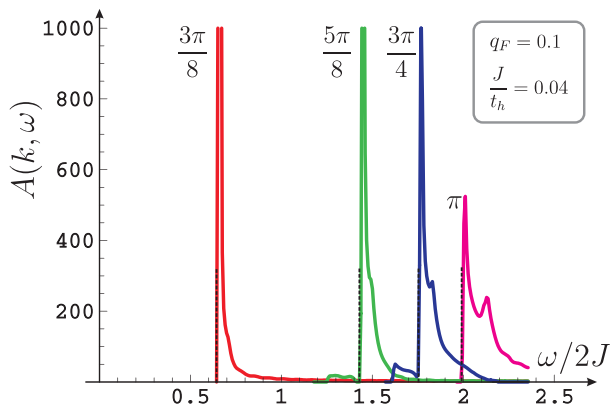}
\caption{Shown is the spectral function $A(k,\omega)$ as a function of $\omega$ for several values of $k$: $3\pi/8$ (red), $5\pi/8$ (green), $3\pi/4$ (blue), and $\pi$ (pink). The curves are obtained from the numerical analysis of the Fredholm determinant in Eq.~\eqref{Afourier}. The positions of the singularities following from the asymptotic expression \eqref{Akomegahresh} are indicated with vertical dotted lines. The parameters of the model used for getting the plot are $Jt_h^{-1}=0.04$ and $q_F=0.1,$ which implies $E_F/2J=0.125.$}
\label{fig:DOSplot}
\end{figure}

To go beyond the asymptotic result \eqref{Akomegahresh} we perform~\cite{zvonarev_bose-hubbard_long08} the numerical analysis of the Fredholm determinant representation of Eq.~\eqref{D}, given in Ref.~\cite{zvonarev_string09}. The resulting plots of $A(k,\omega)$ for $Jt_h^{-1}=0.04$ and $q_F=0.1$ are given in Fig.~\ref{fig:DOSplot}.
One can clearly see in this figure that (i) The position of the threshold singularity is described by Eq.~\eqref{Akomegahresh}. (ii) $A(k,\omega)$ extends below the threshold, which is particularly well seen for $k=3\pi/4.$ This effect is due to the emission by a spin excitation of multiple particle-hole pairs with the dispersion lying below $E_s(k/\nu).$ Within the numerical precision we observe an abrupt cutoff of the spectral weight at some finite frequency below $E_s(k/\nu)$ (for example, at $\omega/2J=1.6$ for $k=3\pi/4$). The reason for such a behavior is unclear. (iii) To the right of the threshold singularity a second peak develops with increasing $k,$ becoming very pronounced for $k=\pi.$ The theoretical explanation of its appearance is an open problem.

In conclusion, we have investigated the dynamics of the spin-$1/2$ Bose-Hubbard model with strong on-site repulsion and small hole doping. We showed that this system exhibits a transition from a Mott-insulator phase to a ferromagnetic liquid. We found collective variables in which spin and charge dynamics separate at {\it all} energy and momentum scales within the $t-J$ approximation. Using these variables we obtain the transverse spin-spin correlation function $G_\perp$, Eq.~\eqref{Gperpdef}, in the form \eqref{GperpH1} and \eqref{D}, convenient for both analytic and numerical investigation. For the spectral function $A(k,\omega)$ associated with $G_\perp,$ we have found the analytic expression for the shape of the magnon peak, Eqs.~\eqref{Akomegahresh} and \eqref{Delta}. Numerical analysis of Eq.~\eqref{Afourier} revealed a non-trivial fine structure of the magnon peak, as shown in Fig.~\ref{fig:DOSplot}.

This work was supported in part by the Swiss National Science Foundation under MaNEP and division II and by ESF under the INSTANS program.


\begin{thebibliography}{99}
\bibitem{rhoapp} In getting Eq.~\eqref{D} we replaced $\varrho_j$ in~Eq.~\eqref{sdec} with its ground state average, which is justified for $1-\nu\ll1.$
\bibitem{giamLL} For the LL formalism we use the notations of Ref.~\cite{giamarchi_book_1d}.
\bibitem{SU2} $SU(2)$ invariance of Eq.~\eqref{Hbose1} implies $\sum_j G_\perp(j,t)=1.$ Using this property one can calculate the prefactor in the asymptotic expression of $G_\perp$.
\end{thebibliography}


\end{document}